\documentclass{evn2002}
\usepackage{graphicx}
\begin{document}
   \title{Higher Resolution VLBI Imaging with Fast Frequency Switching}

   \author{E. Middelberg,\inst{1}
           A. L. Roy,\inst{1}
           R. C. Walker,\inst{2}
           H. Falcke\inst{1}
           \and
	   T. P. Krichbaum\inst{1}
          }

   \institute{Max-Planck-Institut f\"ur Radioastronomie, Auf dem H\"ugel 69, 53121 Bonn, Germany
         \and
             National Radio Astronomy Observatory, P.O. Box 0, Socorro, NM 87801, USA
             }

   \abstract{ Millimetre-VLBI is an important tool in AGN
   astrophysics, but it is limited by short atmospheric coherence
   times and poor receiver and antenna performance. We demonstrate a
   new kind of phase referencing for the VLBA, enabling us to increase
   the sensitivity in mm-VLBI by an order of magnitude. If a source is
   observed in short cycles between the target frequency, $\nu_{\rm
   t}$, and a reference frequency, $\nu_{\rm ref}$, the $\nu_{\rm t}$
   data can be calibrated using scaled-up phase solutions from
   self-calibration at $\nu_{\rm ref}$. We have demonstrated the phase
   transfer on 3C~279, where we were able to make an 86~GHz image with
   $90~\%$ coherence compared to self-calibration at $\nu_{\rm t}$. We
   have detected M81, our science target in this project, at 86~GHz
   using the same technique. We describe scheduling strategy and data
   reduction. The main impacts of fast frequency switching are the
   ability to image some of the nearest, but relatively weak AGN cores
   with unprecedented high angular resolution and to phase-reference
   the $\nu_{\rm t}$ data to the $\nu_{\rm ref}$ core position,
   enabling the detection of possible core shifts in jets due to
   optical depth effects. This ability will yield important
   constraints on jet properties and might be able to discriminate
   between the two competing emission models of Blandford-K\"onigl
   jets and spherical advection-dominated accretion flows (ADAFs) in
   low-luminosity AGNs.}

   \authorrunning{Middelberg et al.}
   \maketitle

\section{Introduction}

The regions where AGN jets are launched and collimated are difficult
to observe with VLBI because the jets are launched very close to the
black hole and most of the bright objects are very distant.  Only in
the closest AGN in Sgr A*, M87, M84, Cen A and M81 can the highest
resolution observations resolve several tens of Schwarzschild radii,
comparable to the scale of 10-1000~$R_{\rm s}$ where jets are
predicted to be launched and collimated (eg Koide et al.  2000, Appl
\& Camenzind 1993). Whilst the collimation regions remain unresolved
in Sgr A* and M81, there are hints for resolving it in M87 (Junor et
al. 1999).

The highest resolution VLBI images are currently achieved from
observations at 86~GHz. However, short atmospheric coherence times and
poor receiver performance and antenna efficiencies severely limit the
observations to sources with $S_{\rm 86\,GHz}\sim0.4~{\rm Jy}$. We
have developed a new phase-referencing strategy for the VLBA that is
not limited by self-calibration at the target frequency $\nu_{\rm t}$,
but only at a lower reference frequency $\nu_{\rm ref}$. A source is
observed switching between these two frequencies on a timescale that
does not exceed half the atmospheric coherence time. After
self-calibrating the $\nu_{\rm ref}$ data, the phase solutions are
multiplied by the frequency ratio $r=\nu_{\rm t}/\nu_{\rm ref}$ and
added to an instrumental, antenna-based phase offset
$\Delta\phi$. Using these phases, the $\nu_{\rm t}$ data can be
imaged.

In a pilot project, we have observed M81 and 3C~279 on January 5,
2002, with various frequency pairs and cycle times to develop the
calibration technique. We describe the details and results of this
project and what we have learnt for future observations.

\section{Observing Strategy and Frequencies}

The Very Long Baseline Array offers the opportunity to switch between
frequencies in less than ten seconds. Frequency switching is done by
moving the subreflector, and therefore the switching time is limited
by its speed (the setup of the electronics only takes one or two
seconds). The frequency selection is governed by three
considerations. 1) The feed horns should lie as close as possible on
the antenna feed circle in the secondary focus to minimize the
switching times and hence the data loss. 2) The phase noise at
$\nu_{\rm t}$ should be as low as possible, and one should therefore
select $\nu_{\rm ref}$ such that the combination of antenna
performance (antenna efficiency and receiver noise) and source flux
density at $\nu_{\rm ref}$ yields little phase noise after
multiplication by $r$. 3) The target source needs to be strong
enough for self-calibration during the envisaged integration time at
$\nu_{\rm ref}$. One should at least expect a $5~\sigma$ detection
within this integration time.

We assumed a nominal coherence time at 86~GHz of 30~s when planning
the experiment and therefore selected an integration time of 15~s per
frequency. We expected a data loss of 5~s due to frequency switching,
leaving 10~s of data at each frequency. Within 10~s, and using
256~Mbps recording for higher sensitivity, we calculated the
theoretical SNRs for three potential reference frequencies 15~GHz,
22~GHz and 43~GHz (Table~\ref{tab:frequencies}). However, the 22~GHz
feed horn has a large angular separation from the 86~GHz feed horn,
requiring long switching times, so we did not consider it further. The
phase noise after scaling by $r$ is about equal for 15 and 43~GHz, and
although the switching time from 43 to 86~GHz is shorter, we preferred
the higher detection SNR at 15~GHz.

\begin{table}
\center
\small
\begin{tabular}{ccccccc}
\hline
$\nu$ & $\theta$ & SEFD & S     & SNR & $r$  & noise\\
 GHz &    deg  &  Jy  &  mJy  &     &      &  deg \\
\hline
15 & 42   & 550   &   169     & 7.8 & 5.61 &  41.2 \\
22 & 102  & 888   &   182     & 5.2 & 3.88 &  42.8 \\
43 & 25   & 1436  &   208     & 3.7 & 2.00 &  31.0 \\
\hline
\end{tabular}
\caption{Possible reference frequencies, the angular separations of
their feed horns from the 86~GHz horn, SEFDs, the expected M81 flux
density as calculated from an estimated 150~mJy at 8.4~GHz (Bietenholz
et al. 2000) and assuming a spectral index of $\alpha=0.2$ (Reuter \&
Lesch 1996), the expected SNR in a 10~s observation, the frequency
ratio to 86~GHz and the expected phase noise after scaling by $r$.}
\label{tab:frequencies}
\end{table}

\section{Observations}

Observations were carried out on January 5, 2002, using all antennas
of the Very Long Baseline Array in excellent weather conditions at all
stations, resulting in atmospheric coherence times of more than 100~s
at 86~GHz. We observed 3C~279 and M81 using all possible frequency
pairs of 15, 43 and 86~GHz, although we did not equally distribute the
observing time among those pairs. M81 was predominantly observed
switching between 15 and 86~GHz because it is at the limit for
self-calibration at 43~GHz. The cycle times on 3C~279 varied from 30~s
(15~s at each frequency) to 90~s, and the cycle times on M81 were 30~s
throughout.

\section{Data Reduction}

\begin{figure}
\centering
\includegraphics[width=0.9\linewidth]{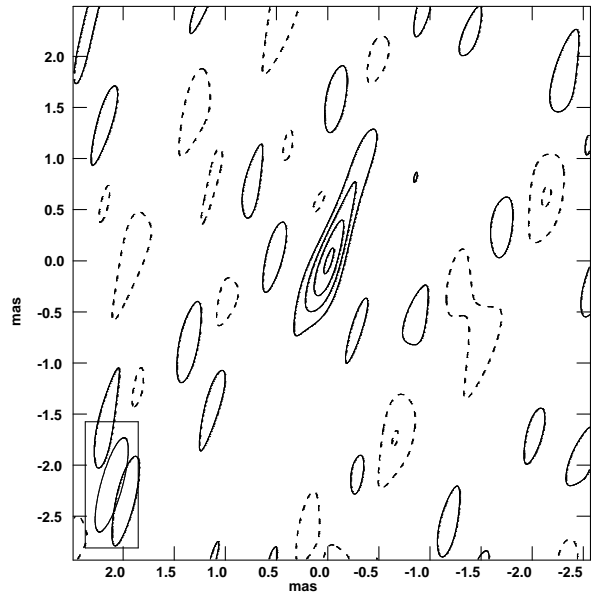}
   \caption{86~GHz image achieved from a 10 minute observation of
   3C~279, switching between 43 and 86~GHz every 30
   seconds. Self-calibration was done with a 10 minute solution
   interval. The peak flux density is 4.17~Jy; contours are drawn
   from  -2 to 4~Jy in steps of 1~Jy.}
      \label{fig:3C279_uncalibrated}
\end{figure}

\begin{figure}
\centering
\includegraphics[width=0.9\linewidth]{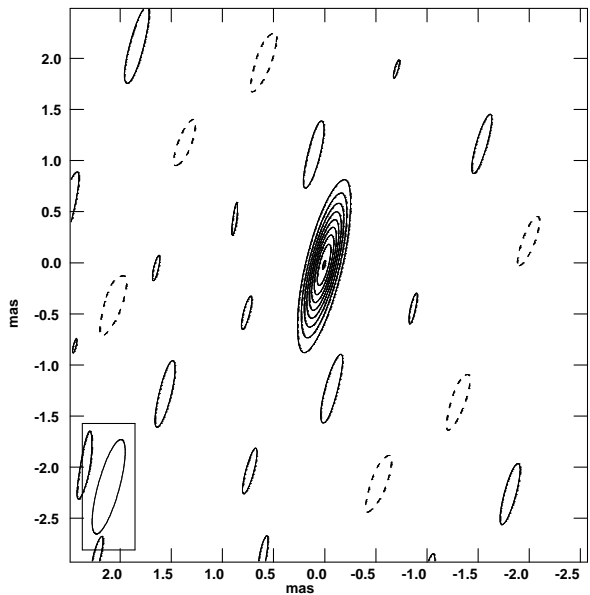}
   \caption{The same data as in Fig. \ref{fig:3C279_uncalibrated}, but
   self-calibration was done at 43~GHz and the phase solutions were
   multiplied by $r$ and transferred to the 86~GHz data. The peak flux
   density has increased to 9.07~Jy; contours are drawn from -2 to
   9~Jy in steps of 1~Jy.}  \label{fig:3C279_phase-transfer}
\end{figure}

\begin{figure}
\centering
\includegraphics[width=0.9\linewidth]{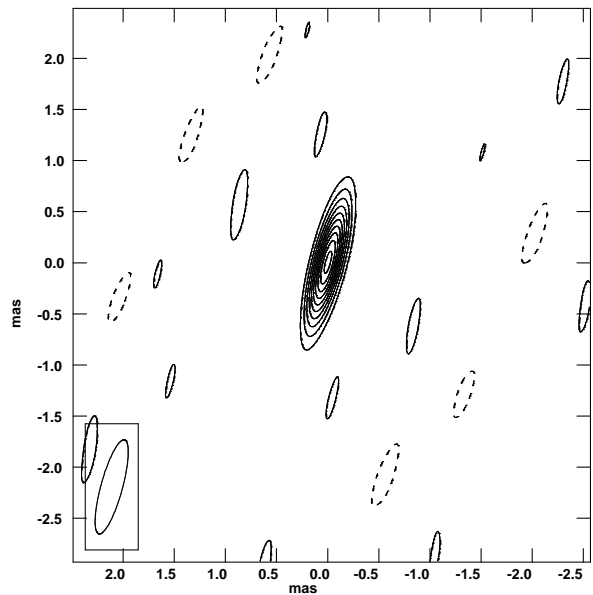}
   \caption{The same data again, but phases were determined using
   self-calibration at 86~GHz with a solution interval of 30~s, equal
   to the scan length. The peak flux density is now 10.4~Jy; contours
   are drawn from -2 to 10~Jy in steps of 1~Jy}
   \label{fig:3C279_fringe-fitted}
\end{figure}

\begin{figure}
\centering
\includegraphics[width=0.9\linewidth]{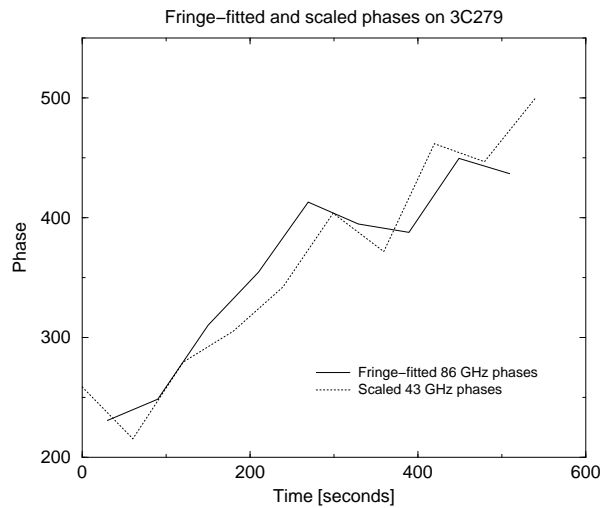}
   \caption{Comparison of the 86~GHz phases (solid line) and the
   scaled 43~GHz phases (dotted line) on the LA-NL baseline. After
   phase wraps have been taken out, the scaled 43~GHz phases follow
   the 86~GHz phases remarkably well.}
      \label{fig:phases}
\end{figure}

\begin{figure}
\centering
\includegraphics[angle=270, width=0.9\linewidth]{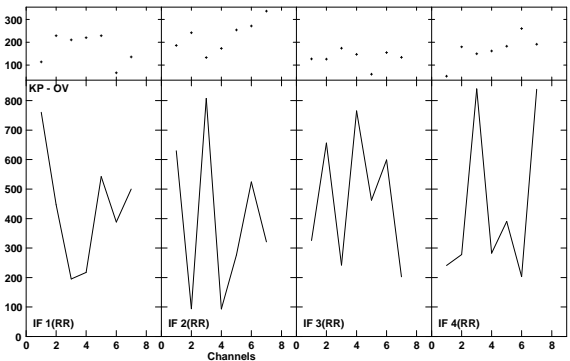}
   \caption{Sample detection of M81 on a short baseline (Kitt Peak to Owens
   Valley) in 20 minutes of data. The top panel shows the phase of the
   vector-averaged cross-power spectrum in degrees, the lower panel
   the amplitude in mJy. Eight channels in each IF have been averaged
   for this figure.}  
\label{fig:KP-OV}
\end{figure}

\begin{figure}
\centering
\includegraphics[angle=270, width=0.9\linewidth]{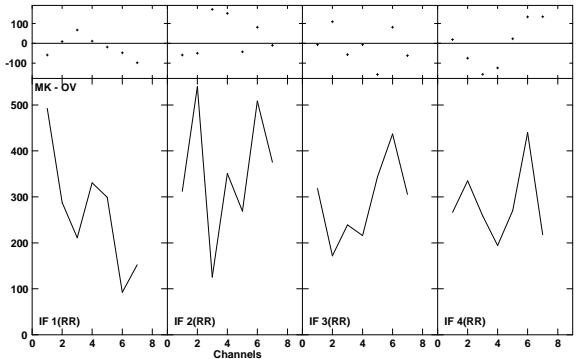}
   \caption{Sample detection of M81 on a long baseline (Mauna Kea to Owens
   Valley) within 20 minutes of data. Axis are the same as in Fig.~\ref{fig:KP-OV}.}
      \label{fig:MK-OV}
\end{figure}

Data reduction was carried out in Bonn, Germany and in Socorro, USA to
exchange experience with the NRAO staff. After loading the data into
AIPS, the 3C~279 data were inspected to develop a flagging scheme
based on the frequency switching times. We found that the first 6 to
7~s of each scan needed flagging, independent of $\nu_{\rm ref}$.  We
noticed that the subreflectors do not always come into position,
resulting in pure noise during the affected scans. The flagging tables
generated by the online systems keep track of those events. We have
merged the online flagging tables with our flagging scheme for optimal
performance. The data were amplitude calibrated using $T_{\rm sys}$
and gain measurements. Fringe-fitting on 3C~279 was performed at each
frequency on a short section of good data at all antennas to
compensate for single-band delays and intra-IF phase offsets. This
allowed averaging across the band during self-calibration in the later
stages of data reduction.

\subsection{3C~279}

We developed the calibration strategy using 10 minutes of data
switching between 43 and 86~GHz (frequency ratio $r=2.00$) every 30~s
on 3C~279. Self-calibration was used to determine the phase solutions
at 43~GHz. The solutions were exported to a plain text file and read
into a Python program to multiply the phase solutions by $r$ and
reformat the data suitable for reading into AIPS.  From each two
adjacent solutions, we calculated the phase rate to allow
interpolation in AIPS. Although for this special frequency pair, $r$
is exactly 2 by good fortune, we soon discovered that the non-integer
frequency ratio $r=5.61$ chosen for the bulk of our M81 observations
caused unpredictable phase jumps after multiplying the phase solutions
by $r$. This can be understood if one pictures a time series of phase
solutions at $\nu_{\rm ref}$, where the phase undergoes a
$360^{\circ}$ turn between two adjacent scans.  This turn of phase
will have no effect at $\nu_{\rm ref}$, but in case $r$ is a
non-integer, the phase at $\nu_{\rm t}$ will show a step, because
$\phi_{\rm ref}\times r~{\rm mod}~360\neq 0$. The problem can only be
solved if one keeps track of all phase turns at $\nu_{\rm ref}$ over
the whole observing run.  We therefore implemented data plotting and
interactive solution editing in our software to remove or introduce
turns of phase in the phase-time series where it seemed obvious.
After scaling the solutions by $r$, the calibrated visibility phases
were constant with time, but an instrumental antenna-based offset
remained. We measured this offset using self-calibration with a
solution interval equal to the length of the observation, in this case
10 minutes.  After this final step, the data were imaged.

The gain in image quality is illustrated in
Fig.~\ref{fig:3C279_uncalibrated}-\ref{fig:3C279_fringe-fitted}. Fig.~\ref{fig:3C279_uncalibrated}
shows the data after self-calibration with a solution interval of 10
minutes. This has brought the visibility phases to zero on average,
but the loss due to atmospheric fluctuations is significant.  Fast
frequency switching (Fig.~\ref{fig:3C279_phase-transfer}) yields an
image that shows $90~\%$ of the flux density shown in
Fig.~\ref{fig:3C279_fringe-fitted}, which has been self-calibrated at
86~GHz.  Thus, the coherence loss is only $10~\%$.
Fig.~\ref{fig:phases} shows the phase solutions obtained via
self-calibration at $\nu_{\rm t}$ and those obtained via
self-calibration at $\nu_{\rm ref}$.

\subsection{M81}

The data reduction was done in the same way as for 3C~279, and
self-calibration with a solution interval of 15~s at 15~GHz yielded
about $97~\%$ good solutions. Due to the non-integer frequency ratios,
it was not possible to integrate longer than 35~min, the longest
continuous time spent cycling between 15 and 86~GHz. Also, only 8~s
out of 30~s delivered good data because we mostly used a short
symmetric cycle time, so the sensitivity was not sufficient after
35~min (i.e. only 9~min of good data at 86~GHz) to determine the phase
offset between 15 and 86~GHz. As a consequence, we were not able to
image M81 and we only made marginal detections in data plots.
(Fig.~\ref{fig:KP-OV} and \ref{fig:MK-OV}).

However, this result is preliminary, and more work is in progress. We
plan to use more sophisticated methods in determining phase solutions
at $\nu_{\rm ref}$, e.g. to consider the ionosphere's electron
content, to estimate the tropospheric delay and to use polynomial fits
to the phase-time series for better interpolation results. Every
single aspect will only have little effect, but together, they might
yield more reliable phase solutions at 86~GHz.

\section{Conclusions}

We have developed a new phase-referencing strategy to make mm-VLBI
observations of sources too weak for self-calibration. The strategy is
based on rapid changes between the target frequency and a reference
frequency and uses scaled-up reference frequency phase solutions to
calibrate the target frequency. We have observed 3C~279 as a strong
test source and M81 as a weak science target. Using this technique, we
were able to image 3C~279 at 86~GHz with a $10~\%$ coherence loss
compared to conventional self-calibration, and, for the first time, we
were able to make a marginal detection of M81 at 86~GHz.

We have learnt three basic lessons from our project: 1) To prevent
loss of coherence occuring from phase wraps at $\nu_{\rm ref}$, one
should select frequencies such that their ratio is an integer. Given
the VLBA frequency agility, one can almost always tune the frequencies
to satisfy this condition. 2) To increase efficiency, one should use
asymmetric switching times. The symmetric times that we used gave us
only 8~s, or $25~\%$ of data per cycle of 30~s duration. For future
projects, we will use cycles of 60~s with 16~s at $\nu_{\rm ref}$, 7~s
for switching, 30~s at $\nu_{\rm t}$ and 7~s for switching back to
$\nu_{\rm ref}$. This is only possible in good weather conditions,
when the atmospheric coherence time is 60~s or longer, but the VLBA's
dynamic scheduling ensures a high probability. 3) One should use
regular scans on a strong calibrator to monitor the phase offset
between the frequencies of interest.

Fast frequency switching turns out to be a powerful tool for highest
resolution VLBA images of weak sources. In some of the nearest, but
unfortunately radio-weak AGNs, it might reveal the jet collimation
region on scales of tens of Schwarzschild radii. However, it can also
be used at lower frequencies, eg. to phase-reference 15~GHz
observations using interleaved 5~GHz scans. Atmospheric coherence and
frequency switching times should be less problematic at these
frequencies.

Fast frequency switching is basically a phase referencing technique,
so it will reference the $\nu_{\rm t}$ data to the core position at
$\nu_{\rm ref}$. After phase transfer, the $\nu_{\rm t}$ phase on a
given baseline will contain a time-dependent offset which comprises
the sum of a constant instrumental term, $\Delta\Phi_{\rm instr}$, and
a term due to the source geometry at $\nu_{\rm t}$, $\Delta\Phi_{\rm
geo}$. If the core at $\nu_{\rm t}$ is shifted with respect to the
$\nu_{\rm ref}$ core, $\Delta\Phi_{\rm geo}$ will be a sinusoid whose
period is 24~h, whose amplitude is a measure for the core shift
between the two frequencies and whose zeros give the direction of the
shift. $\Delta\Phi_{\rm instr}$ and $\Delta\Phi_{\rm geo}$ can be
separated by monitoring $\Delta\Phi_{\rm instr}$ on a calibrator and
fitting a sinusoid to the residual phase solutions on the source.

Using this technique, it should be possible to look for core shifts
due to optical depth effects in radio jets, because the bulk of
emission in jets comes from the $\tau=1$ surface, and this surface is
expected to move closer to the central mass when higher frequencies
are observed. In low-luminosity AGNs like Sgr~A* and M81, two emission
models have been suggested, using either scaled-down jets like in
quasars (Falcke et al.  1993), or spherical advection-dominated
accretion flows (ADAFs; Rees 1982; Melia 1994; Narayan et
al. 1995). These models can be distinguished because only the jet-like
emission should have a core shift, and we have proposed to look for it
in Sgr~A* and M81 using a variety of frequencies.

\begin{acknowledgements}
The VLBA is an instrument of the National Radio Astronomy Observatory,
a facility of the National Science Foundation, operated under
cooperative agreement by Associated Universities, Inc. EM acknowledges
partial support from the EC ICN RadioNET (Contract
No. HPRI-CT-1999-40003).
\end{acknowledgements}

\end{document}